# On the Relative Role of East and West Pacific Sea Surface Temperature (SST) Gradients in the Prediction Skill of Central Pacific NINO3.4 SST

Lekshmi S[1,] Rajib Chattopadhyay*[1,2,] D.S. Pai[3], M. Rajeevan[4], Vinu Valsala[2], K.S. Hosalikar[1], M. Mohapatra[5]

1. India Meteorological Department, Pune-411005
2. Indian Institute of Tropical Meteorology, Pune-411008
3. Institute for Climate Change Studies, Kottayam, Kerala-686004
4. National Centre for Earth Science Studies, Thiruvananthapuram, Kerala-695011
5. India Meteorological Department, New Delhi-110003

*Corresponding Author email: rajib@tropmet.res.in

# Abstract

Dominant modes of SST in the west and east Pacific show strong but regionally different gradients caused by waves, internal dynamics, and anthropogenic warming, which drives air-sea interaction in the Pacific. The study discusses the relative contribution of SST gradients over the western and eastern Pacific to the prediction skill of SST in the central Pacific, where El-Nino, La-Nina, or El-Nino Modoki events project significantly. For this, the analysis develops a convolutional neural network (CNN) based prediction model to predict the Nino3.4 SST. CNN-based prediction models use a spatial filter at the initial stage, which is highly efficient in capturing the edges or gradients and hence are useful to understand the role of SST spatial gradients in the prediction skill. The study reports three CNN-based model experiments. The first one is a CTRL experiment that uses the whole equatorial Pacific domain SST pattern. The second and third models use the equatorial eastern and western Pacific domain SST only. Another novel feature of this study is that we have generated a large number of ensemble members (5000) through random initialization of CNN filters. It is found that random initialization affects the forecast skill, and the probability density function of the correlation skill of the 5000 models at each lead time shows a gaussian distribution.

The model experiments suggest that the west Pacific SST model provides better Nino3.4 skills as compared to the east Pacific skill. The CNN-based model forecast based on the SST pattern, thus, shows the impact of the SST spatial pattern on the ENSO forecast.

# 1. Introduction

The central pacific sea surface temperature (SST) prediction is essential from the perspective of El-Nino Southern Oscillation (ENSO) prediction, especially the El-Nino Modoki events (Ashok et al., 2007), CP-El-Nino (Yang and Jiang, 2014) and understanding teleconnections (e.g. ENSO-Monsoon predictive teleconnection (Sahoo and Yadav, 2021) and several times dynamical models miss reliable prediction of central pacific SST (Wen et al., 2021). Different flavors of ENSO, as well as El-Nino Modoki events, all project on the central Pacific SST with different intensities. The spatial patterns of sea surface temperature (SST) over the equatorial central Pacific are controlled by local air-sea interaction (i.e., SST-convection feedback)(Zhang et al., 1995; Xiang et al., 2013), ocean advection(Kug et al., 2009), wind-induced thermocline variation (Ashok et al., 2007), subtropical atmospheric forcing(Yu et al., 2010) and the delayed effects of the wave propagation (Suarez and Schopf, 1988). The current prediction skill of SST or ENSO is determined by the efficiency of models in representing these physical and dynamical processes in the western, eastern, and central Pacific Ocean. The thermocline is shallow, and SST is colder on the east Pacific. The SST is warmer, and the thermocline is deeper in the western Pacific. Over the colder SST, the convective heating will be less effective, and hence, air-sea interaction is weaker, resulting in weak feedback to the atmospheric process contributing to SST variability. This happens over the eastern Pacific for example (refer to Fig.1 of Suarez and Schopf, (1988) for example). On the western side, the situation is different. The thermocline is much deeper, and again coupling is weaker like the eastern Pacific. However, westward propagating Rossby wave reflection occurs in the western boundary, and the reflected Kelvin wave propagates eastward. This is able to produce stronger temperature change (i.e., dT/dt) in a delayed mode in the central Pacific and determines the local SST pattern. How does the spatial distribution of equatorial SST in the eastern and western Pacific Oceans contribute to the prediction skill of equatorial SST or ENSO? It is well known that the ENSO or SST prediction skill has seasonality, and the spring barrier in predictability is well-studied (Torrence and Webster, 1998; Mukhin et al., 2021).

The importance of SST gradients in air-sea interaction and convection is well known (e.g. (Lindzen and Nigam, 1987). Understanding the dependence of the prediction skill of central Pacific SST or ENSO on the spatial temperature pattern of the equatorial Pacific is, thus, important (Liang et al., 2021). Central Pacific El-Nino and El-Nino Modoki have shown an increasing trend in recent years (Ashok et al., 2007). As discussed earlier, the SST evolution pattern over the eastern and the western pacific are dominated by different processes, and the depth of thermocline differs largely over the eastern and western Pacific, we undertake the current study to show the relative prediction skill of central Pacific SST as well as ENSO by using the SST pattern over the western and eastern Pacific. How does the spatial temperature pattern (or SST gradient) over the Pacific can impact the prediction skill? It can be answered in two ways: *first*, the dispersion relation connects the spatial structure of a mode with its temporal frequency. Rossby mode has a distinct spatial structure which is related to its periodicity, and hence, temporal evolution tendencies of this mode are linked to the spatial pattern that impacts the prediction skill. **Second**, the advection term in the temperature equation is related to the spatial gradient of a field, which leads to contributing to the calculation of temperature tendencies (e.g. equation 1 of (Ballester et al., 2016). The advective terms in the ocean determine low-frequency oscillations of general ocean-atmosphere interaction(Saravanan and McWilliams, 1998) and the ENSO advective oscillators (Picaut et al., 1997). The dynamical models can be used to understand the budget terms, and contribution of different physical terms can be calculated (Wen et al., 2021). However, the SST variation in dynamical models depends on a large number of physical processes, and the sensitivity in predicting SST anomalies are dependent on multiple physical factors. Often isolation of simpler factors requires well-designed experiments. Even the simplest oscillator models (Zebiak and Cane, 1987) recognize the importance of SST variation in the eastern and western Pacific but do not consider spatial gradients. For example, the west pacific oscillator (Weisberg and Wang, 1997), central and east Pacific oscillators (Zebiak and Cane, 1987) are somewhat different in conceptual formulation those describe the SST variations in the western and eastern side of the basin.

Thus, the above studies suggest an important role of spatial SST patterns in the context of understanding the ENSO. However, asymmetries in the processes that contribute to ENSO dynamics are well known, and the recent studies based on AI-based methods clearly brought out such features (Shin et al., 2022). In the context of the prediction of central Pacific SST, the role of west pacific versus east Pacific surface temperature patterns are not clear. More specifically, how the advective gradients, or the wave patterns that form the SST structure in the east and west Pacific, contribute to the prediction skill is still not clear. Global warming or other factors affect surface temperatures. West and East Pacific warming trend is different owing to different physical factors (Park et al., 2022), and hence warming trend would impose different gradient of temporal evolution on these sectors. Hence, an estimation of the sensitivity of surface temperature patterns on the prediction skill is important.

Deep Machine learning models are increasingly used in prediction problems, and they are now used in the ENSO prediction problem which suggest skill varying from 6 months to 18 months (Mu et al., 2021; Ham et al., 2019; Guo et al., 2020). Deep learning models are very efficient in isolating spatial gradients, and the training does not suffer from vanishing gradient problems as compared to normal fully connected networks. Different variants of convolutional Neural Network (CNN) based deep learning models are popular in the ENSO prediction. In the context of understanding the role of west Pacific and east Pacific pattern in predicting the central Pacific SST (or El-Nino Modoki and CP-El-Nino), the CNN based model can be used. The CNN models have spatial filters and pooling layers that identify the dominant spatial gradients in the data. Thus, the spatial sample is denoised keeping the most important information. A prediction scheme using the CNN thus can be suitable for understanding the role of SST spatial patterns in predicting the ENSO.

The current analysis will frame a simple (i.e., "vanilla") CNN based prediction model of central Pacific SST. Unlike other models those try to extend the prediction skill of ENSO indices to longer duration by combining several steps, this *vanilla* CNN model is used here to understand the role of SST spatial pattern in the west and east Pacific to predict the central pacific SST and hence CP-ElNino or ElNino Modoki. Such simple models are based on basic recommended AI model version and are conveniently used to understand dynamics of ENSO events (Shin et al., 2022). The prediction model is formulated based on traditional CNN based method and uses one variable only (i.e., SST). The prediction experiment attempts to predict NINO3.4 SST anomaly, which is considered as an important central Pacific indicator of ENSO. The schematic of a CNN is shown in **Fig.1.** Additionally, in the current analysis, we will generate a multi-ensemble prediction models using a CNN formulation, to construct a probabilistic ensemble of CNN based prediction models and we show the relative role of west and east Pacific SST in the prediction of NINO3.4 SST at different lead times.

## 2. Method and Data

(a) *CNN based prediction scheme*

The CNN model used here is developed by using SST over the Pacific domain as the predictor. The method closely follows Ham et al (2019) except the domain is different and the input matrix also uses SST lags up to 12 month as different channels. The reason for selection of a lag of twelve months is as follows: we first conducted a simple forecast using multiple linear regression using NINO3.4 as predictor and predict the same index at different lead-times. We appended multiple lags to generate maximum skillful prediction of NINO3.4 at lag 1 and then the correlation skill decreases as lead increases (**Fig.7** black curve, also discussed latter). This is taken as a benchmark for the CNN forecast CTRL experiment (discussed in next section). The same lag is then used for CNN prediction with the hypothesis that at higher lags the CNN should at least provide better skill. For the CNN model also, the predictand is the area averaged SST anomaly over the NINO3.4 region (area-averaged SST anomaly over 170°–120° W, 5° S–5° N) up to 12 months lead-time.

(b) *Prediction Experiments*

To understand the relative role of the SST over west and East Pacific, three experiments are designed. The first experiment uses SST distribution over the whole equatorial Pacific (120° E-90° W,15° S-15° N) which is used as a standard reference of *CONTROL* experiment in which the convolution operation is performed over the whole Pacific domain and the convolution filters are applied for the whole domain. The domain selected shows strong gradients in trend and SST pattern can be obtained from empirical orthogonal function analysis of SST. The dominant trend mode and the second dominant mode shows strong gradients (plot for trend and EOF patterns are not shown, refer Barbosa and Anderson (2009) or L'Heareux et al., (2013) for detailed analysis. Also, since the full Pacific domain is selected, this model is supposed to have maximum information for the convolution operation. This is referred as the *CTRL* forecast through out this paper. In the second experiment CNN model is created based on the eastern Pacific (160° W-90° W,15° S-15° N) SST anomaly and is referred to as the *EAST* experiment. This model has only east Pacific SST gradient information. The third experiment takes only west pacific domain (120° E-160° W,15° S-15° N) and is referred as *WEST* experiments. This model has only west Pacific SST gradient information.

We have created 5000 ensemble experiments for each of the three types of runs with skin temperature data and 1000-member ensemble for ERSST and COBE SST runs. The last two analysis datasets are primarily used for verification (discussed in next section). The ensemble forecast is created to understand the sensitivity of the stochastic randomization of the filter weights. Ensembles are generated in such a way that different filter weights are generated and stored after the model fit. Each of the ensemble member models are stored separately. It is found that such weight randomization affects the predictability skill in the CNN models. This is analogous to the sensitivity of dynamical model runs to slight variations in initial conditions (i.e., effect of chaos and non-linearity) or effect of displacement error while computations causing forecast drifts (Orrell et al., 2001). Model is trained from 1950-2001 (612 months) and test period is 2002-2021 (228 months).

The block diagram of the three experiments are shown in **Fig.1**. As already mentioned, the SST conditions for the past 12 months have been used to predict the NINO3.4 SST anomaly conditions for a lead time of 12 months. The basic components for all the three models remain the same with the input domain size varying for different experiments. The model consists of two convolutional layers, two pooling layers, a Flatten layer and a Dense layer. The input is filtered using the convolution layer with 64 filters each having a kernel size of 3 X 3. The output from the first convolution layer undergoes a spatial pooling by using a 2 X 2 max-pooling layer with a stride of 2. This filtered data is again passed through a convolution layer having 128 filters with a kernel size of 3 X 3 and a max-pooling layer same as the above. This is flattened using a Flatten layer and the output is obtained by passing through a Dense layer of size 12. Both the convolution layers use hyperbolic tangent function (tanh) as the activation function and the Dense layer uses a linear activation function. The initial weights for the Convolution layers are created using the Glorot Uniform (also known as Xavier Uniform) initializer with the seed taken as 10 and the biases are initialized with zeros. A summary of the set of hyperparameters used for the model runs are shown in **Table 1**. The models are run using the tensorflow keras python module in the computing resources provided by the Google Colaboratory which includes the NVIDIA-System Management Interface (SMI) driver version of 460.32.03.

(c) Data

The CNN network requires data without undefined value. To fulfill the objective of the analysis we will use surface temperature data. There are several sea-surface temperature reanalysis data are available. However, sea surface temperature (SST) data generally contains undefined value over land and small islands. These undefined values are not desired in the convolutional network routines. Hence, we will use skin temperature as our primary data and sea-surface temperature data (ERSST V5(Huang et al., 2017) and COBE SST (Hirahara et al., 2013) with temperature filled over land as secondary verification data for our conclusion. The monthly SST data is obtained from the daily skin temperature data from NCEP Reanalysis 1 dataset for a period from 1950-2021 (Kalnay et al., 1996) over the tropical Pacific domain (120° E-90° W,15° S-15° N). The skin temperature data represents the SST over oceans and the land surface temperature over the continents. This data is highly correlated (> 0.85) with ERSST analysis and COBE SST analysis over equatorial pacific (plot not shown). The monthly skin temperature data is calculated from the daily data and the anomaly is derived by removing the climatological average during 1950-2021. The SST anomaly for the NINO3.4 region which is the predictand, is obtained by taking the arithmetic average of the skin temperature anomaly over the NINO3.4 region (170°–120° W, 5° S–5° N). The spatial skin temperature anomaly data (hereafter mentioned as SST) for the past 12 months are used as the predictor to predict the NINO3.4 SST anomaly for a lead time of 12 months.

## 3. Results

The current analysis is based on the assumed relation of the east and west Pacific with central Pacific. To have a first order understanding of the same relationship, we plot the monthly lag-correlation of SST averaged over a small reference box (170°W - 140° W, 5°S-5°N) in the central Pacific with the entire equatorial pacific between 15°S-15°N. The lag-correlation is shown in **Fig.2**. The correlation analysis shows that there is a systematic shift of higher correlation pattern from lag-12 to lag-1 indicating that at higher lags west Pacific is correlated to the box. As lag decreases, east Pacific gets higher correlation. Thus, it indicates the shift in the spatial correlation pattern with respect to lags. This systematic shift in the spatial pattern of correlation for central Pacific SST indicates that to develop a model for SST prediction in the central Pacific, both east and west Pacific Oceans are important. To understand the relative importance of the west and east Pacific SST patterns we perform the experiments as described in sec.2. The west and east Pacific domains chosen for experiments are shown in **Fig.1 (top).**

a. *CTRL experiment*: The CTRL experiment spans the whole equatorial Pacific domain (refer sec.2b). The **Fig.3a** (left panel) shows the training and test loss for this experiment with the shaded portion showing the spread among ensembles. Model shows a realistic stabilization in the training and test mode. The skill of the 5000-member ensemble up to 12 months lead-time is shown in **Fig.4.** The correlation coefficient /standardized RMSE (normalized by predictand standard deviation during the test period) is plotted along the x-axis, while the percentage distribution of ensemble is plotted along the y-axis. Each subpanel represents a lead-time with skills plotted up to 12 months lead-time. Correlation stays in the range 0.5-0.6 and normalized (standardized) RMSE also stays below 1 for all members up to 6 months lead-time. The mean absolute error (MAE), which is used as the loss function to configure the model is also shown in the figure. MAE shows increase similar to RMSE but it is starting from some lower value from lead-1 to lead-8. Also, the spread of error distribution increase with lead time. Thus, the model gives good fidelity in NINO3.4 forecast up to 6 months. It may be noted that as the lead increases correlation spread among members increases while the RMSE spread decreases. This correlation skill is comparable to dynamical model skill (Barnston et al., 2011, 2012). The skill, however is modest as compared to the current generation of advanced CNN model with multiple predictors or bigger domains e.g. (Mu et al., 2021; Ham et al., 2019). However, the purpose of the experiment is to see the sensitivity of west Pacific versus East Pacific SST pattern in determining the skill. Hence, we will use this as a baseline skill.

b. *EAST experiment*: For the EAST experiment the skill is shown in **Fig.5** in a similar manner as that of **Fig.4.** The training and test loss are shown in middle panel of **Fig.3b**.

For the east pacific experiment, the skill quickly falls below 0.5 at 5 months lead time and also, the RMSE increases at a faster rate with respect to lead-time. The spread (i.e. half width) of the distribution is larger than the CTRL experiment. MAE plot also shows similar result as RMSE plot.

  c. *WEST experiment:* The plot for WEST experiments are shown in a similar manner in **Fig. 6** and the training and test loss are shown in right panel of **Fig.3c**. As it can be seen that in comparison with EAST experiment the WEST experiment shows much better skill and the skill is comparable to the CTRL prediction. Also, the spread is comparable to the CTRL (i.e. full domain). The MAE distribution plot also shows lesser spread than the EAST experiment.

The above results are summarized in terms of ensemble mean skill in **Fig.7**. The results suggest that the West Pacific experiment shows better results in term of correlation and RMSE. At the shorter lead-time (2-5 months) West Pacific provides higher skill. **Fig.7** also shows the prediction skill of NINO3.4 based on a regression-based prediction model developed using NINO3.4 SST up to 12 lags as predictors (a multilinear regression using lags as different variables for the same training period). This is the benchmark model to verify the CTRL run. As it can be seen that though at lower lead (lead-time 1 or 2) the skill of CTRL CNN model and the benchmark model is comparable, at the higher lead-time the CNN based model give better skill (lead-time 6 above), confirming the usefulness of CNN based machine learning forecasting. The power spectra of forecasted SST and observed SST at higher lead-time shows that the power is weaker in the forecast at higher lead-time (not shown). This indicates that forecasted variance is biased, which is a common problem in forecast which require variance correction (which we have not attempted here though, as our aim is to compare the sensitivity of skill with east/west domain). The temporal evolution of the observed NINO3.4

SST anomaly versus predicted anomaly (without any bias correction) for all the experiments as different lead time forecasts during the test period is shown in **Fig.8.** It is clear from the plot that the models have difficulty in capturing the 2015-2016 event extreme peak as the lead-time increases. Also, for higher lead-time some phase shift can be noticed (e.g. lead 5 for the year 2011). To understand the skill during peak EL-Nino or La-Nina, the model forecasts for the four peak El-Nino and La-Nina events are shown in **Fig. 9** and **Fig.10** respectively. The results show that at higher lead-time the skill of the forecasts from WEST runs are better on many occasions.

## 4. Comparison with other Reanalysis Data

In order to understand the fidelity of the results based on skin temperature data, we have compared the results with runs from two separate analysis data (ERSST and COBE SST analysis). The ERSST and COBE SST is used to create a 1000-member ensemble. The comparison is made for 1000-member ensembles of all the three datasets. The results of prediction are shown in **Fig.11.** From **Fig.11b** and **11c** it is clear that like the skin temperature (Fig.**11a**), the WEST runs have better correlation than EAST runs. Also, the CTRL shows better correlation at six months lead time for NCEP skin temperature and reanalysis but not in ERSST where the correlation is not significant. Thus, in general the results indicate that EAST and WEST basin shows different skills at different lead-times.

## 5. Discussion and Conclusion

The above analysis is carried out to compare the NINO3.4 forecast skill based on WEST and EAST runs and compare with a whole domain (CTRL) run. The primary experiments are conducted using the NCEP skin temperature and not with ERSST or COBE SST analysis as the SST has undefined data over islands in the domain. Analysis SST data is used as additional source of verification in which the SST over the island is replaced by NCEP skin temperature. The objective behind this comparison is to see the relative role of western Pacific and eastern Pacific surface temperature pattern in the prediction skill of the NINO3.4 SST which is a representative index for different ENSO flavors including ElNino Modoki patterns (e.g. Table-2 of (Ashok et al., 2007). The SST patterns are reflected as the horizontal spatial gradients which are resultant of trends, physical processes, advective properties or wave modes. Such SST gradients are well known to impact the local convection and surface winds(Lindzen and Nigam, 1987) . Hence the pattern of SST gradients can impact the prediction skill. The current study employs an efficient deep machine learning based gradient detection method (CNN) to predict the SST index (NINO3.4) in the central Pacific. The CNN use the convolution operations and the convolution filter identifies the gradients (of SST in the current study) in the spatial domains. The results clearly indicate that West Pacific SST pattern plays a dominant role in determining the forecast skill of NINO3.4 SST. The skill is highest for the whole domain (i.e., CTRL) run when the model has full information of the equatorial SST pattern in the Pacific. But the skill is modest when only central to eastern Pacific patterns are considered. The asymmetric impact (i.e. WEST has more skill than EAST) reflects complexity of ENSO dynamics in the Pacific, as discussed in other literatures which use machine learning to understand the role of ENSO dynamics. The prediction skill is comparable to the dynamical model forecast for the CTRL runs Barnston et al., 2011, 2012), thus indicating that the surface SST in the western equatorial Pacific plays a major role in the predictability of NINO3.4 pattern. Western and eastern Pacific have asymmetric trend in terms of warming, and physical processes that govern the surface as well as subsurface SST distribution is different. Also, central pacific SST is impacted by the Rossby waves and the reflected Kelvin waves from the western Pacific boundary, and the thermocline is deeper in the west. All these factors determine the surface SST pattern. We show here that the surface SST pattern of western side and eastern side plays a different role in the CNN models which employ the spatial filters to generate the prediction model at the initial layers.

Another important conclusion from the study is that, the initialization of spatial filters in the CNN layers matter in determining the prediction skill. Even a small difference weight of the filters (due to randomization at the model fit layer) can produce large difference in skill among the ensemble members at larger lead-time. Thus, it is important to take care of the randomization of the filters in understanding the prediction skills. It is probably best to create a large number of ensemble and the ensemble distribution would be given to fully understand the forecast efficiency of a machine learning model.

The difference in skill related to WEST and EAST runs can be attributed to the fact that the ENSO phases and their reversal are related to the delayed Oscillator response of Rossby wave in the western Pacific, which changes the phases of the ENSO after Rossby reflection in the western boundary as Kelvin waves. The reflection of Kelvin waves in the eastern Pacific as coastal Kelvin wave across the eastern coast, probably does not help in the ENSO oscillatory cycle as much as in the WEST runs. For the peak ENSO cases (ElNino or LaNina) considered here, the WEST run shows moderately better SST values than EAST runs. Skill for some individual El-Nino or La-Nina cases are comparable in the EAST, WEST or CTRL runs indicating that the peak ENSO forecast skill may or may not depend on SST of the western and eastern Pacific oceans. Subsurface information is not included here and its role on ENSO prediction skill would be seen in a later study.

## Acknowledgements

LS acknowledge the research fellowship support from MRFP Project, Ministry of Earth Sciences (MoES), Govt of India. Authors acknowledge, Head CRS (IMD) for the support. Research support from Indian Institute of Tropical Meteorology (IITM), an autonomous institute under MoES, and India Meteorological Department (IMD) is acknowledged. Authors also acknowledge DGM, IMD for discussions and support.

## Data availability statement

The primary skin temperature, ERSST V5 and COBE SST data used here are freely available and is mentioned in the text. Any derived data/code will be made available on a reasonable request.

## Conflict of Interest

The authors declare no conflict of interest.

**Figure Captions**

**Figure 1**: A schematic diagram of the CNN model used. The three experiments (CTRL, EAST (EP), and WEST(WP)) are conducted using similar configurations but differ only in the domain size. N denotes the sample size. For each SST spatial pattern sample of the past 12 months lag, the model outputs Nino3.4 SST anomaly for the next 12 months lead.

**Figure 2**: Spatial pattern of the correlation coefficient of Pacific SST with respect to area-averaged SST taken from the small reference box (170°W - 140° W, 5°S-5°N) in the central Pacific. The correlation pattern shows the asymmetric pattern of response and shift in different lags.

**Figure 3**: The training and test loss as a function of epochs for the **(a)** CTRL, **(b)**EAST and **(c)** WEST experiments.

**Figure 4**: The verification metrics (correlation, normalized(standardized) RMSE and mean absolute error (MAE)) as a function of lead time for the test period (2001-2021) for the CTRL experiment. Each row in the plot represent one lead-time mentioned at the right side of each panel. The probability distribution (shaded) is based on the 5000 ensembles. For the domain size of CTRL experiment refer Figure 1.

**Figure 5**: Same as Figure 4 but for EAST experiments.

**Figure 6**: Same as Figure 4 but for WEST experiments.

**Figure 7**: The verification metrics (correlation and Standardized RMSE) as a function of lead-time for the ensemble mean for all the experiments. The metrices are plotted along the ordinates.

**Figure 8**: Temporal evolution of the observed SST anomaly (°C) versus the predicted SST anomaly at different lead-times for all the experiments during the test period from 2002-2021.

**Figure 9**: Skill of the model prediction at different lead time during the El-Nino peak conditions that occurred during the test period of 2002-2020. The observed Nino3.4 SST anomaly (°C) and the forecasted SST for the different runs at different lags up to 6 months are shown as different bars.

**Figure 10**: Same as figure 9 but for La-Nina conditions.

**Figure 11:** Comparison of the correlation and RMSE of CTRL, WEST and EAST Experiments trained using different SST data such as **(a)** NCEP Skin Temperature **(b)** ERSST V5 SST **(c)** COBE SST. A mean of 1000 ensembles each has been used for comparison for the same period from 2002-2021.

Table 1: Summary of the Model Configuration

| Model Type | Convolutional Neural Network (CNN) |
|---|---|
| **Learning rate** | 0.001 |
| **Epochs** | 150 |
| **Batch size** | 4 |
| **Loss Function** | Mean Absolute Error (MAE) |
| **Optimizer** | RMSprop |

# FIGURES

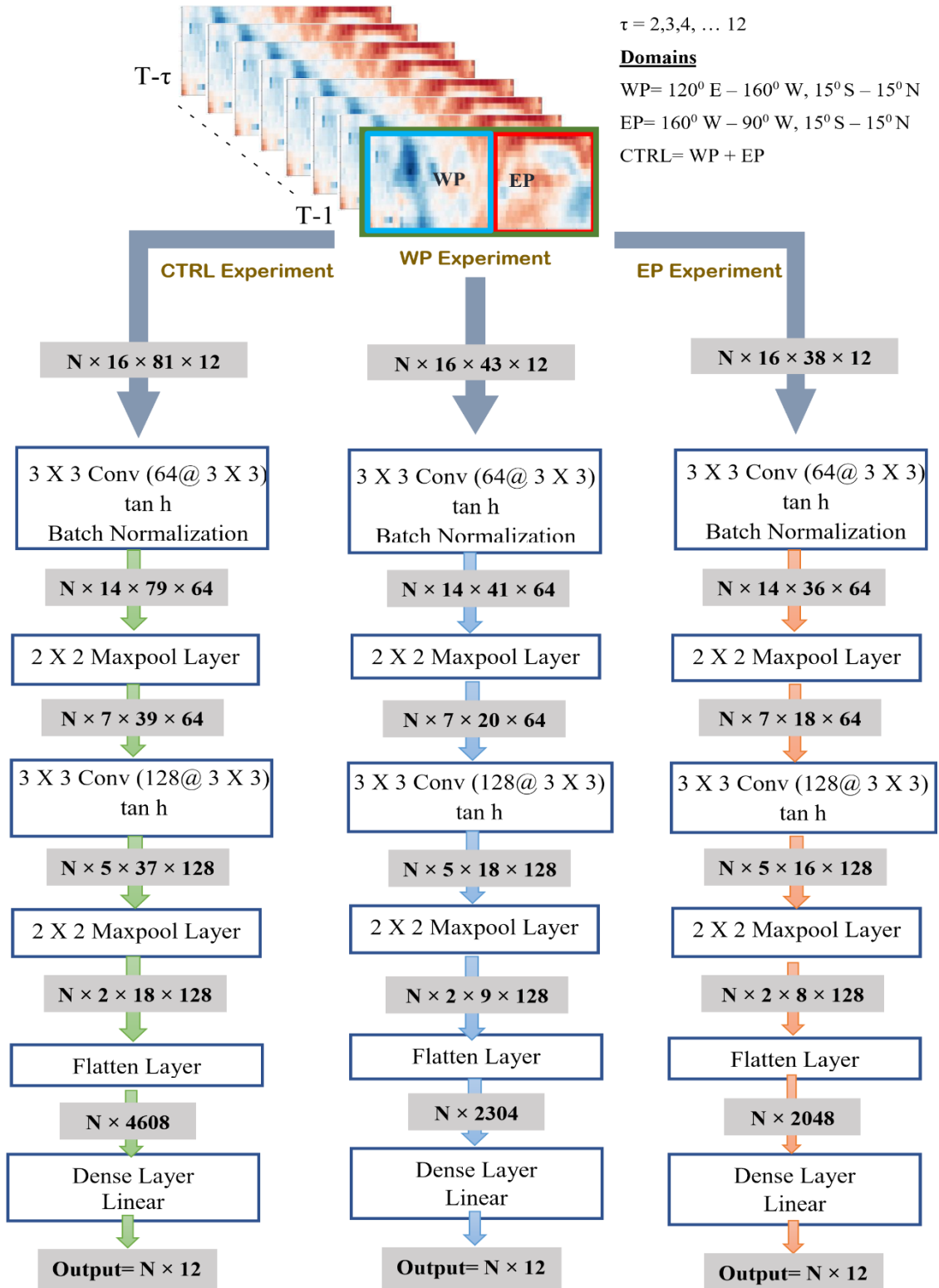

*Figure 1*: A schematic diagram of the CNN model used. The three experiments (CTRL, EAST (EP), and WEST(WP)) are conducted using similar configurations but differ only in the domain size. N denotes the sample size. For each SST spatial pattern sample of the past 12 months lag, the model outputs Nino3.4 SST anomaly for the next 12 months lead.

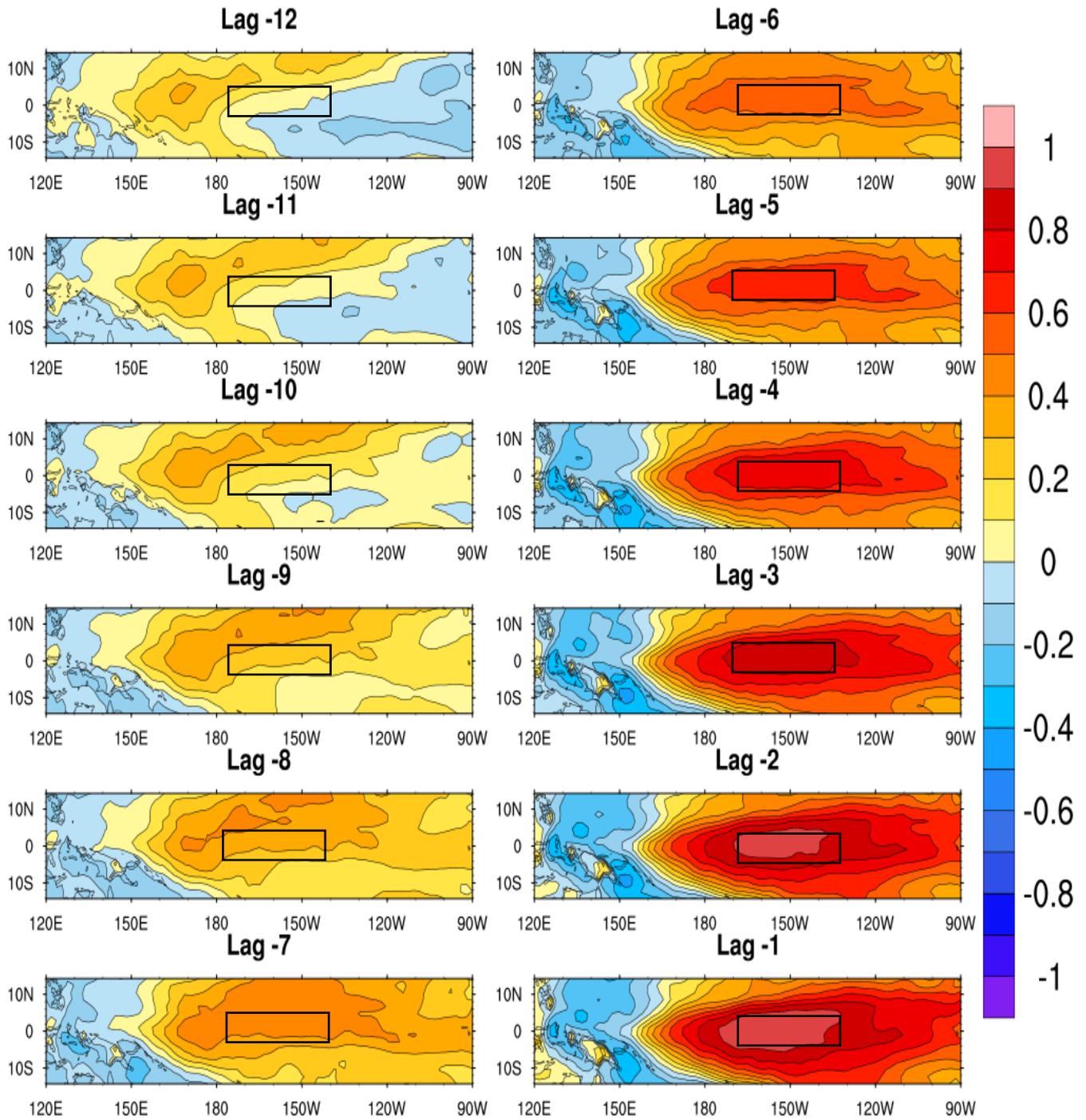

*Figure 2*: *Spatial pattern of the correlation coefficient of Pacific SST with respect to area-averaged SST taken from the small reference box (170°W - 140° W, 5°S-5°N) in the central Pacific. The correlation pattern shows the asymmetric pattern of response and shift in different lags.*

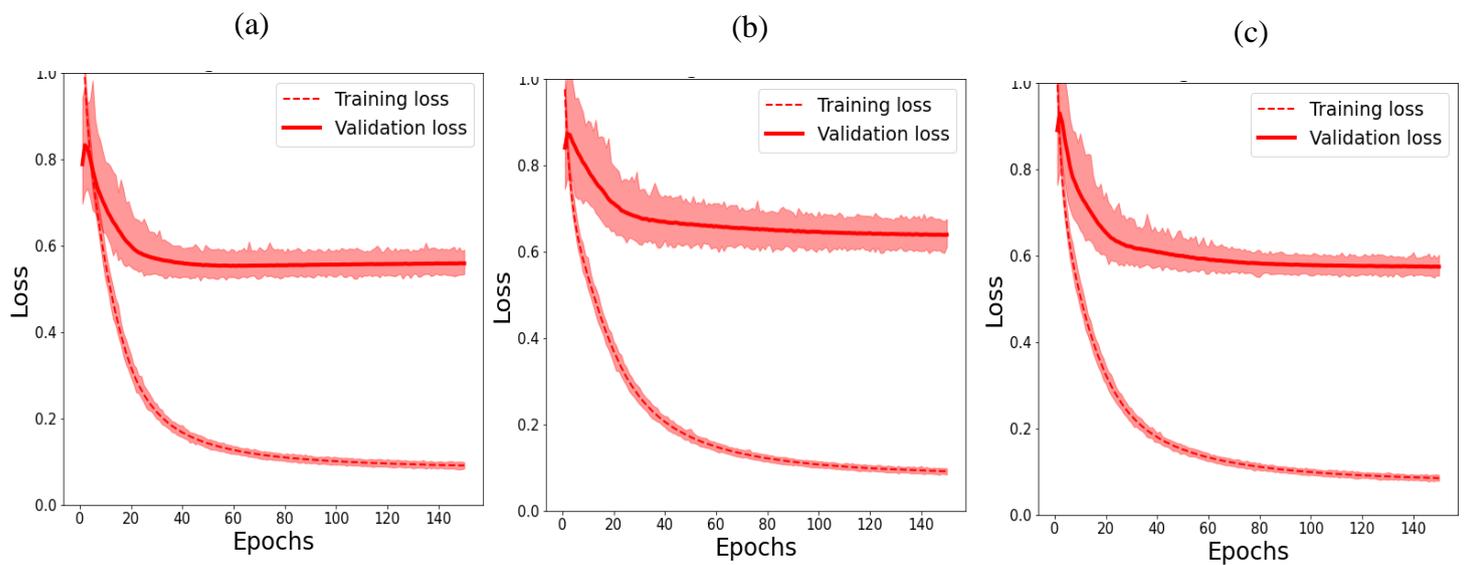

*Figure 3*: *The training and test loss as a function of epochs for the (**a**) CTRL, (**b**)EAST and (**c**) WEST experiments.*

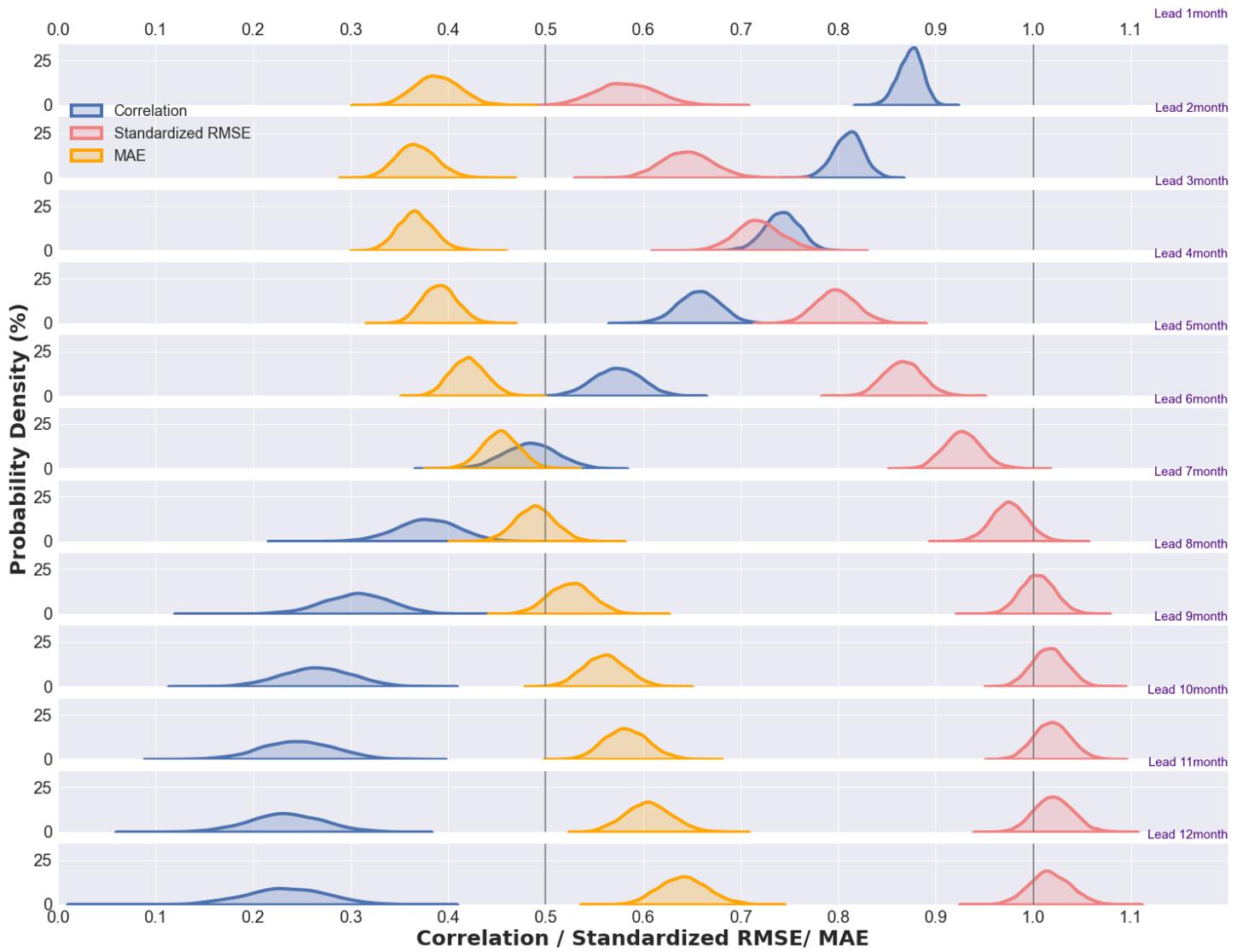

*Figure 4*: *The verification metrices (correlation, normalized(standardized) RMSE and mean absolute error (MAE)) as a function of lead time for the test period (2001-2021) for the CTRL experiment. Each row in the plot represent one lead-time mentioned at the right side of each panel. The probability distribution (shaded) is based on the 5000 ensembles. For the domain size of CTRL experiment refer **Figure 1**.*

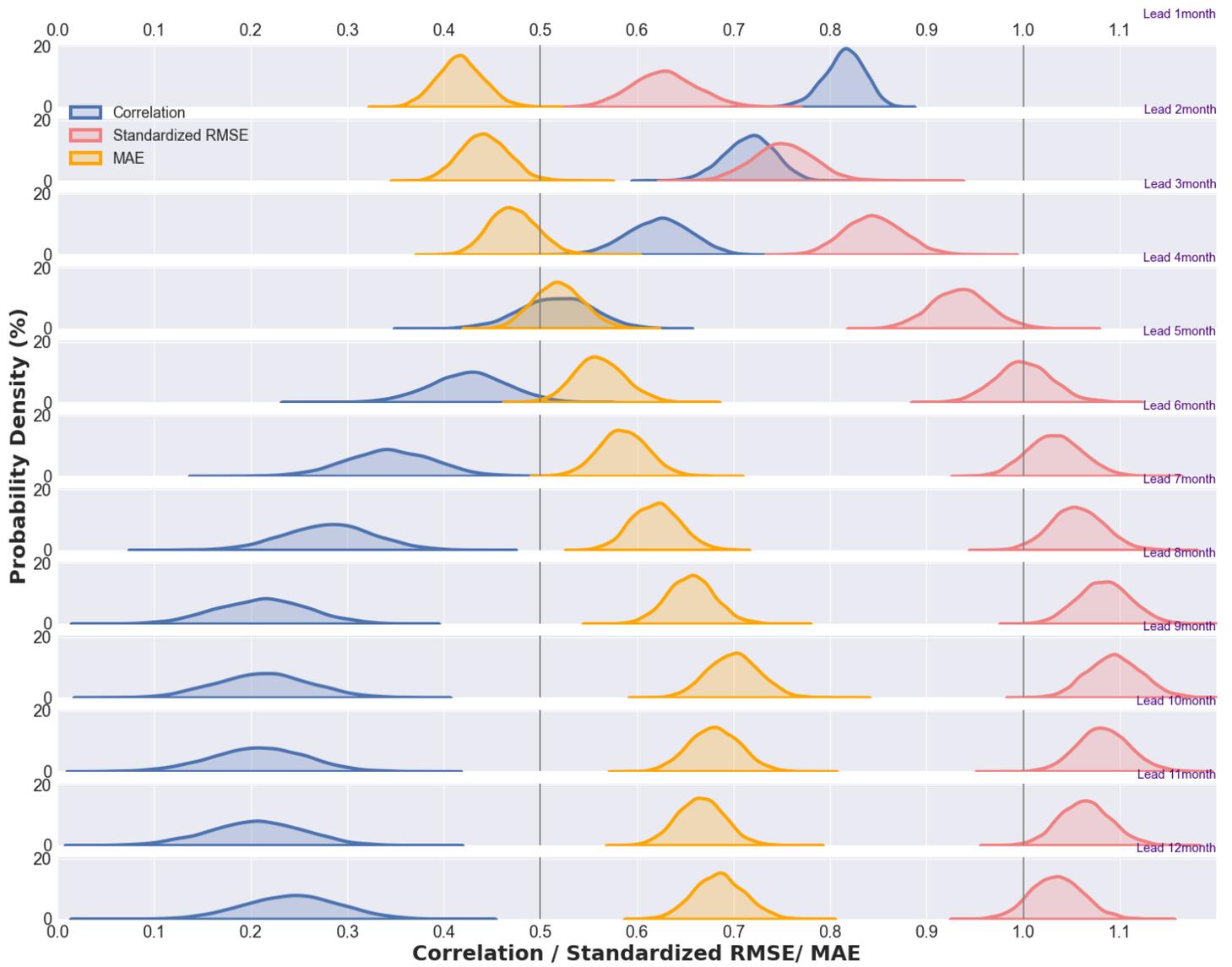

*Figure 5*: Same as *Figure 4* but for EAST experiments.

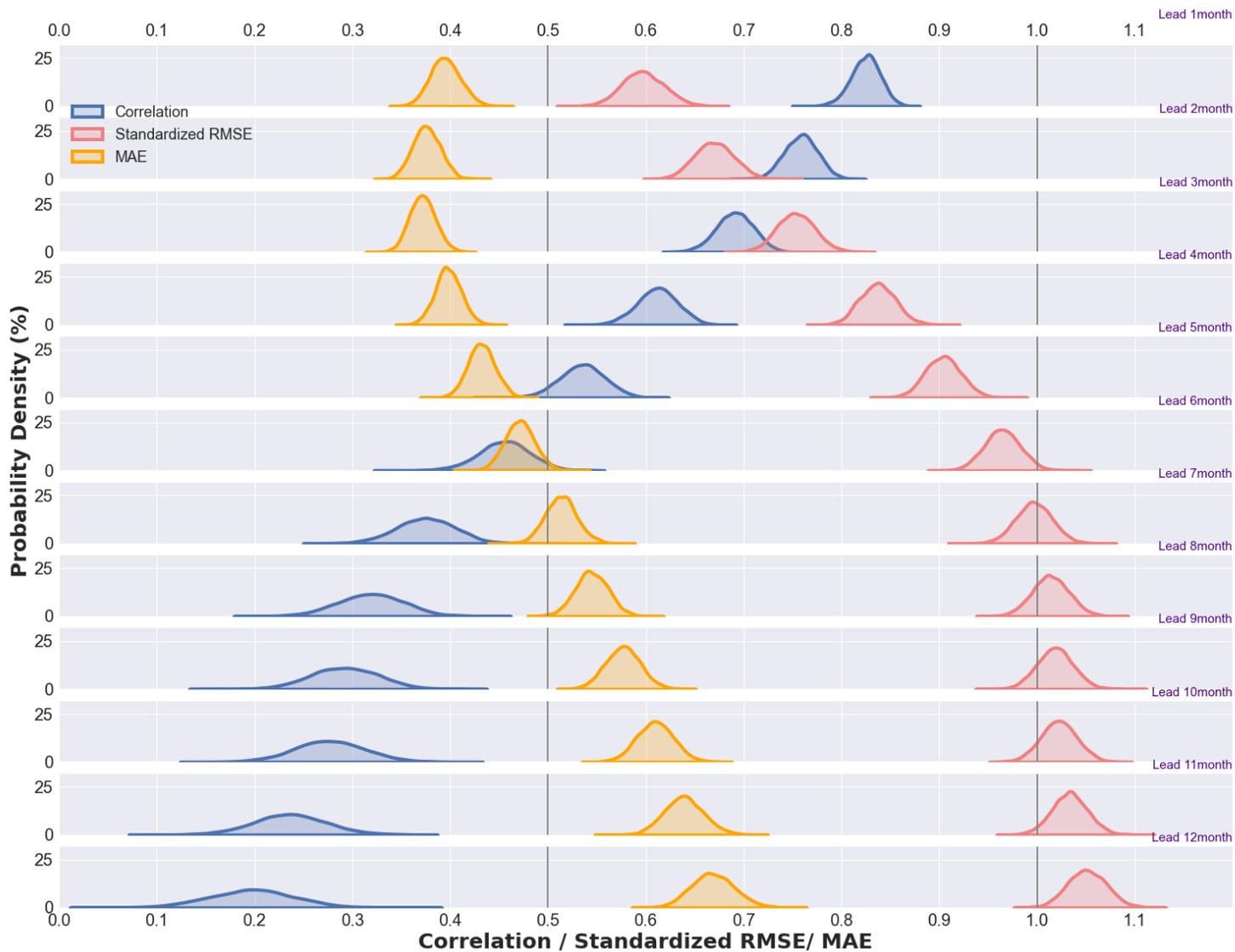

*Figure 6*: Same as *Figure 4* but for WEST experiments.

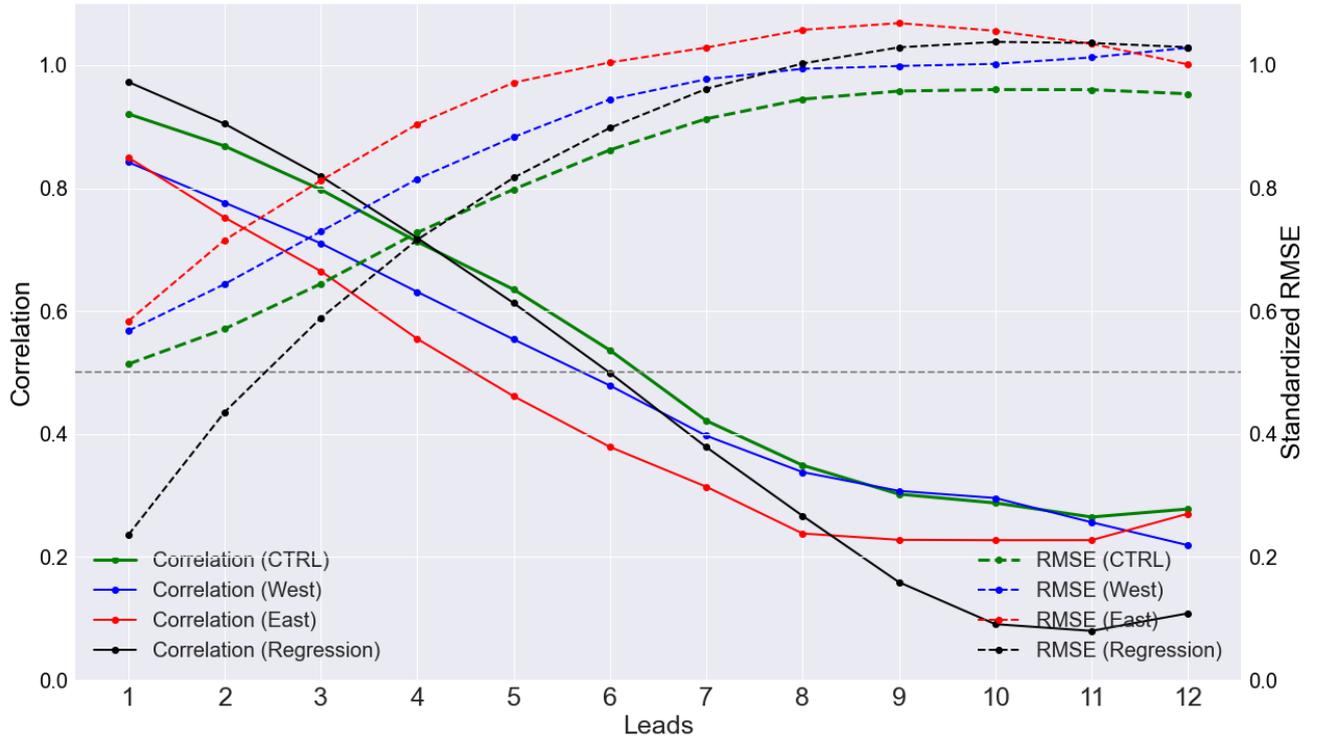

***Figure 7****: The verification metrices (correlation and Standardized RMSE) as a function of lead-time for the ensemble mean for all the experiments. The metrices are plotted along the ordinates.*

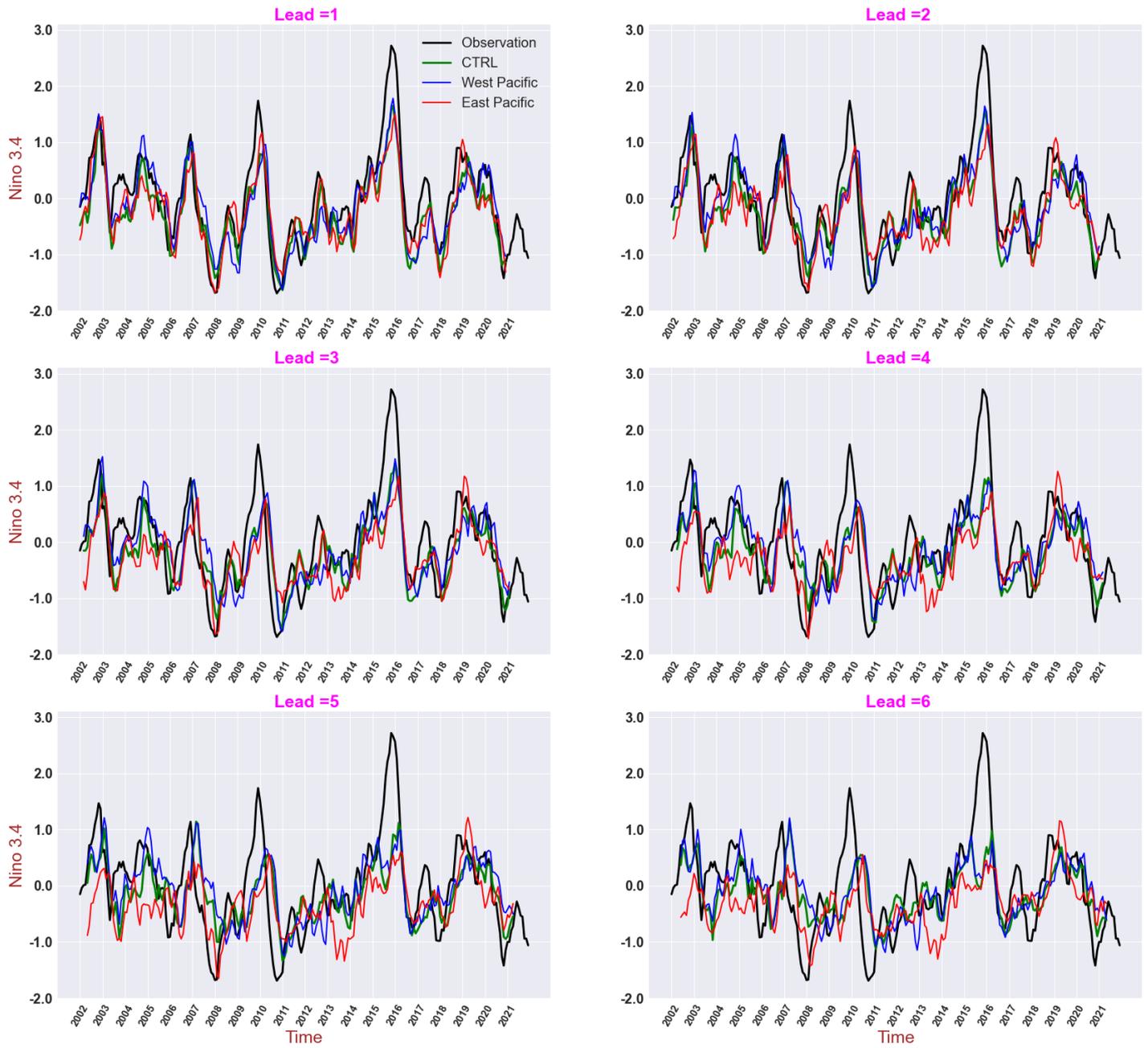

***Figure 8****: Temporal evolution of the observed SST anomaly (°C) versus the predicted SST anomaly at different lead-times for all the experiments during the test period from 2002-2021.*

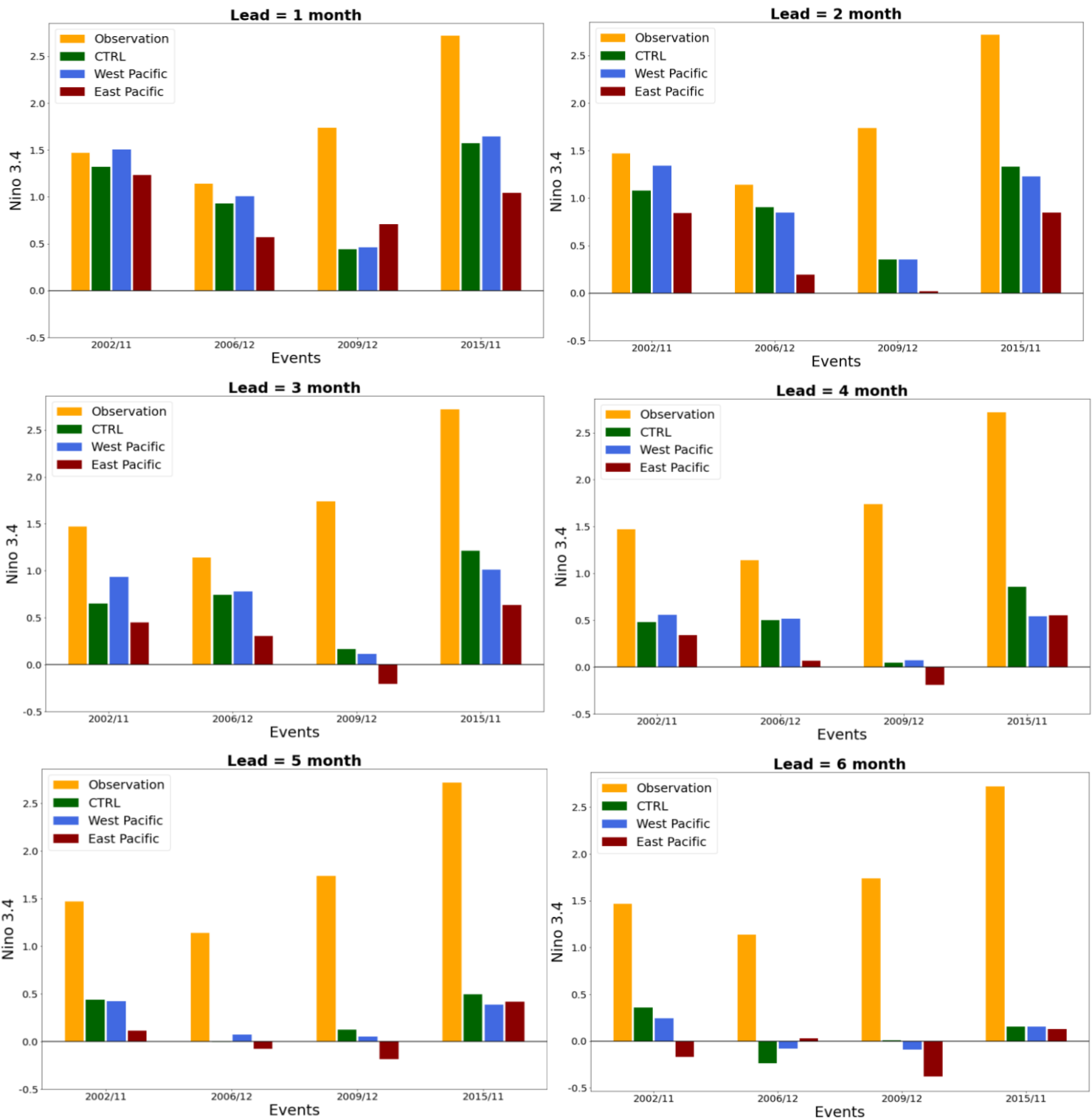

***Figure 9***: *Skill of the model prediction at different lead time during the El-Nino peak conditions that occurred during the test period of 2002-2020. The observed Nino3.4 SST anomaly (°C) and the forecasted SST for the different runs at different lags up to 6 months are shown as different bars.*

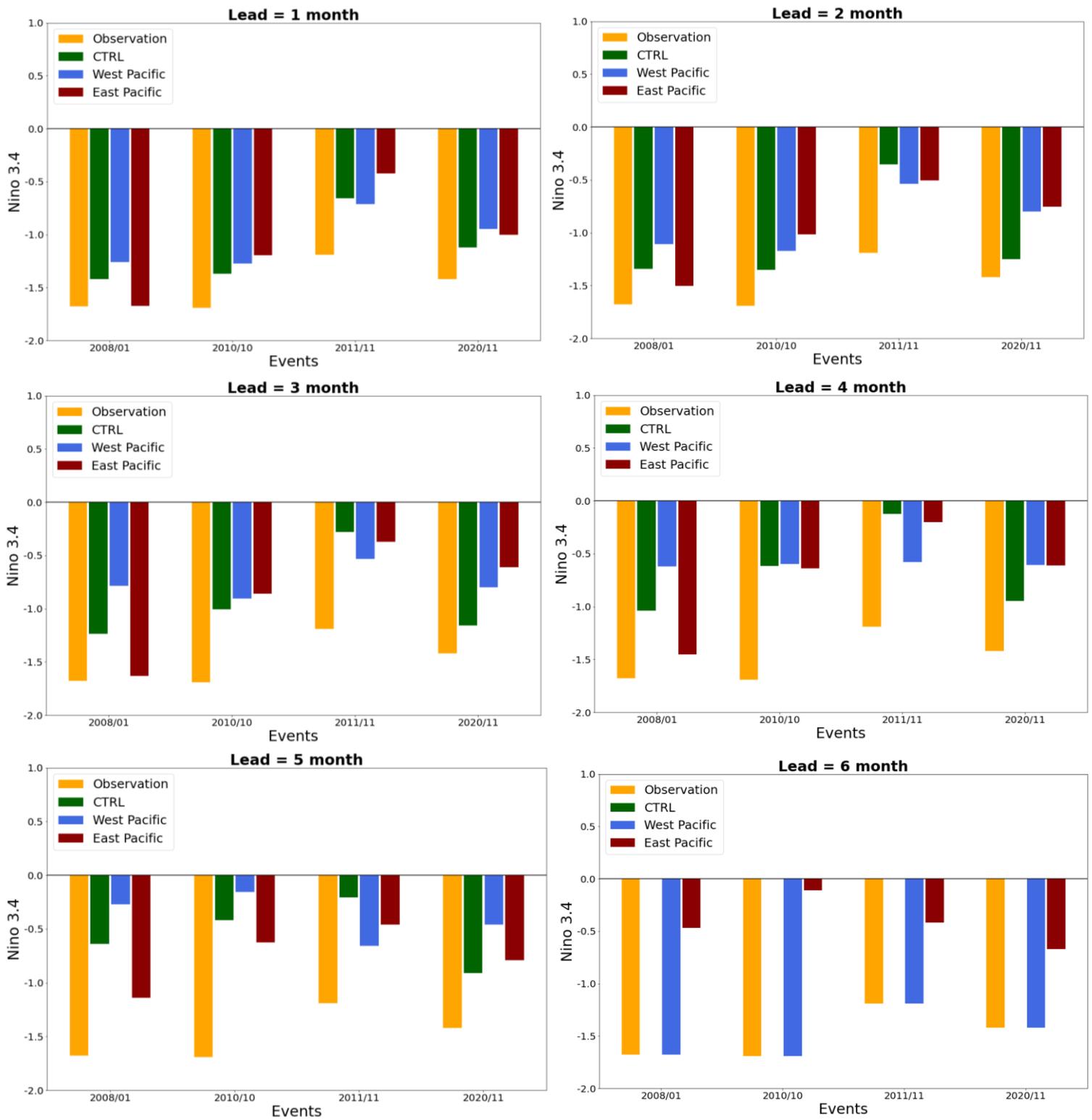

*Figure 10: Same as **figure 9** but for La-Nina conditions.*

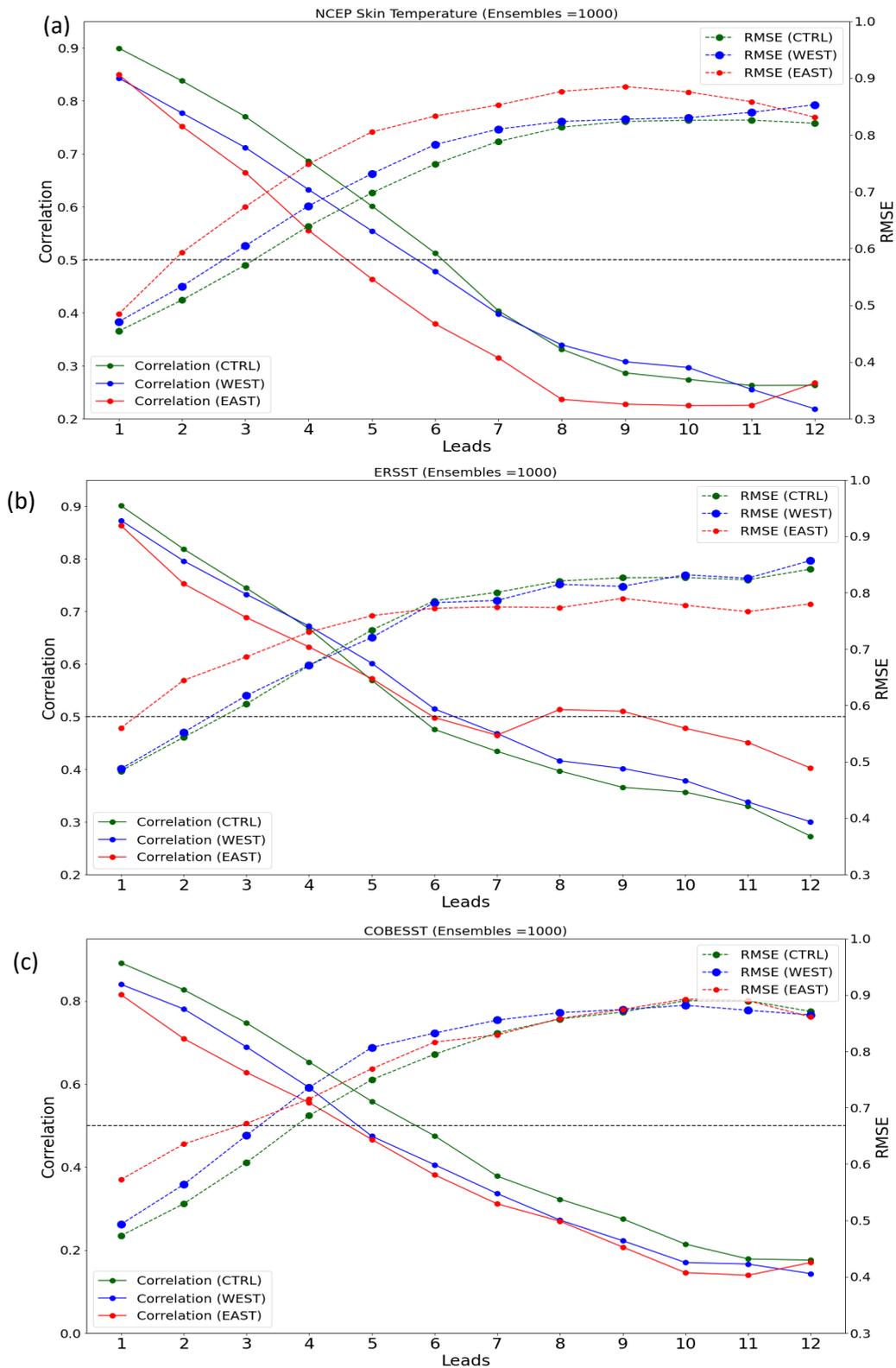

***Figure 11***: *Comparison of the correlation and RMSE of CTRL, WEST and EAST Experiments trained using different SST data such as (**a**) NCEP Skin Temperature (**b**) ERSST V5 SST (**c**) COBE SST. A mean of 1000 ensembles each has been used for comparison for the same period from 2002-2021.*